# Seasonal modulation of seismicity: the competing/collaborative effect of the snow and ice load on the lithosphere

Antonella Peresan [1,5], Francesco Cocetta [2], Giuliano F. Panza[3,4,5]

[1] National Institute of Oceanography and Experimental Geophysics, CRS-OGS, Trieste. Italy.
[2] University of Trieste, Department of Mathematics and Geosciences. Trieste, Italy
[3] Institute of Geophysics, China Earthquake Administration, Beijing, People's Republic of China
[4] Accademia Nazionale Lincei - Rome. Italy
[5] ISSO. International Seismic Safety Organization, Arsita, Italy

**Abstract**

Seasonal patterns associated with stress modulation, as evidenced by earthquake occurrence, have been detected in regions characterized by present day mountain building and glacial retreat in the Northern Hemisphere. In the Himalaya and the Alps, seismicity is peaking in spring and summer; opposite behaviour is observed in the Apennines. This diametrical behaviour, confirmed by recent strong earthquakes, well correlates with the dominant tectonic regime: peak in spring and summer in shortening areas, peak in fall and winter in extensional areas.

The analysis of the seasonal effect is extended to several shortening (e.g. Zagros and Caucasus) and extensional regions, and counter-examples from regions where no seasonal modulation is expected (e.g. Tropical Atlantic Ridge). This study generalizes to different seismotectonic settings the early observations made about short-term (seasonal) and long-term (secular) modulation of seismicity and confirms, with some statistical significance, that snow and ice thaw may cause crustal deformations that modulate the occurrence of major earthquakes.

# 1. Introduction

Tectonic forces responsible for mountain building must overcome, among others, gravity force (Doglioni et al., 2015); this fact may give rise to competing effects between tectonic forces and the load due to snow and ice cover formation. Heki (2003) and Panza et al. (2011) evidence that snow and ice loads acting on mountain building of Japan, Alps and Himalaya cause seasonal crustal deformation and perturb the inter-seismic strain buildup. Earthquakes in these regions occur on reverse faults, the snow load enhances compression at reverse faults, reducing the Coulomb failure stress by a few kPa, a value sufficient to modulate the tectonic stress buildup (few tens of kPa/yr). The increment of the vertical load in a region of tectonic shortening increases the vertical while decreases the deviatoric stress components; consequently, the shear stress remains below the critical value and the faults are more "stable". Accordingly, the number of strong earthquakes, that we identify here and in the following with the events with $M_i \gtrsim M = M_{max} - 1.5$ ($M_{max}$ being the maximum recorded magnitude in each study area) decreases when the snow load is present. While in shortening regions covered with snow in winter, inland earthquakes tend to concentrate in spring and summer, the earthquakes in extensional regions (Apennines) tend to concentrate in fall and winter (Panza et al., 2011). The recent strong earthquakes that occurred in the study areas support these findings: in the Himalaya, where the seismicity is peaking in spring and summer, the Nepal event of April 25, 2015 fits the pattern. In the Apennines, where opposite behaviour – peak in fall and winter – is found, the cluster of strong earthquakes that struck Central Italy starting on August 28, 2016 and still active in winter 2017, also supports this pattern. Although the statistical significance of this seasonal correlation is not strong, due to the limited number of studied regions and to the time span of available observations, these preliminary results point towards the seasonal modulation of seismicity that is controlled by the regional tectonic regime.

A quantitative analysis of seasonal patterns associated with meteorological stress modulation and earthquake occurrence has been published in several papers in the past few years (Bettinelli et al., 2008; Bollinger et al., 2007, Christiansen et al., 2005, 2007; Gao et al., 2000; Grapenthin et al., 2006; Heki, 2003; Saar and Manga, 2003). Snow and ice load, with respect to other seasonal meteorological phenomena (e.g. rainfalls), is characterized by much longer residence time and relatively more continuous and homogeneous distribution over the surface of the orogeny.

In this paper, using earthquake catalogues fully reliable for our purposes, we generalize and confirm, with some statistical significance, the early observations made about short-term (seasonal) and long-term (secular) modulation of seismicity, considering the study areas shown in Fig. 1.

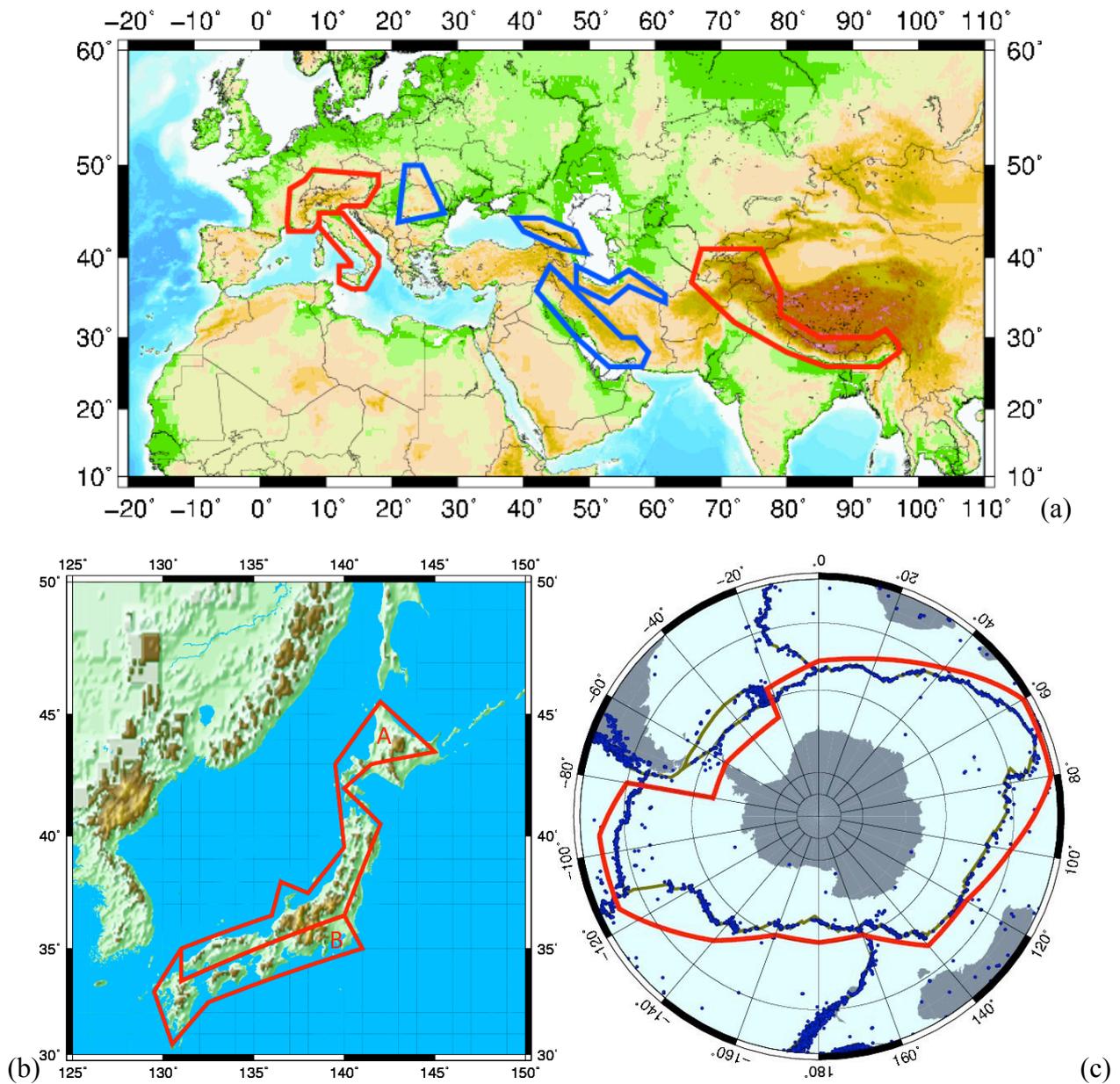

*Fig. 1 – Polygons defining the investigated areas: a) and b) represent Northern Hemisphere study regions, whereas c) represents Antartica plate that is completely circumscribed by spreading centers or "ridges": the Circum-Antarctic region (dots represent epicentres recorded in the period 1990-2013). In map a) regions previously studied by Panza et al. (2011), i.e. Alps, Apennines, Himalaya, are contoured in red, while in blue are contoured the regions studied for the first time in this paper, i.e. Vrancea, Caucasus, Zagros, Elburz. In map b) letter A identifies the Japanese territory usually affected by snow falls and letter B the territory with lack of snow falls.*

## 2. Seasonal effect in Antarctic region

Antarctica is a lithospheric plate completely circumscribed by spreading centres and it experiences a well known seasonal change in the snow/ice load. As suggested by the theoretical computations of Heki (2003), the analysis of the seasonal effect is performed considering only earthquakes occurred within the crust. In the study area, suitable data are provided by global NEIC catalogue (NEIC-USGS, 1989; ANSS, 2017) for the time interval 1990-2013, during which, for all events, the origin time is specified, at least, to the day. Following Panza et al. (2011), as a rule, the earthquakes with $M_i \geq M=M_{max}-1.5$ are considered, thus essentially accounting for events that involve the entire crust (Heki, 2003). The available data obviously do not permit any investigation of long-term (secular) effect.

As measure of seismicity, we consider the number of events ($N$) and the normalized strain ($\Sigma$) released by them. $\Sigma$ is computed using Benioff strain release $S_i$ (Benioff, 1951), calculated for each earthquake i with magnitude Mi, and normalized to the strain $S_{min}$ of the minimum magnitude, $M_{min}$, of the earthquakes considered for the analysis, that is:

$$\Sigma = \sum_i \frac{S_i}{S_{min}} = \sum_i 10^{\frac{d}{2}(M_i - M_{min})} \qquad (1)$$

where $d=1.5$ is used as in Gutenberg and Richter (1956).

In Fig. 2 the earthquakes along the Circum-Antarctic extensional region have been conventionally grouped accordingly to the Southern Hemisphere meteorological seasons: winter (WI: June, July and August), spring (SP: September, October and November), summer (SU: December, January and February), fall (AU: March, April and May). The seasonal (fall-winter) peak is clearly visible in Figs 2a and 2b and corroborates Panza et al. (2011) findings for the Apennines (extensional region). To account for the variability of meteorological seasons (climatologic seasons do not always correspond to astronomic ones) the beginning of each season is arbitrarily shifted up to one month (e.g. austral summer starting on December 1st, 15th and 31st, respectively - the remaining seasons being shifted accordingly): the seasonal effect evidenced is stable and it remains stable even if magnitude threshold is slightly lowered (Fig. 3). Therefore, the result obtained in the Circum-Antarctic region is fully consistent with Panza et al. (2011) findings for the Apennines.

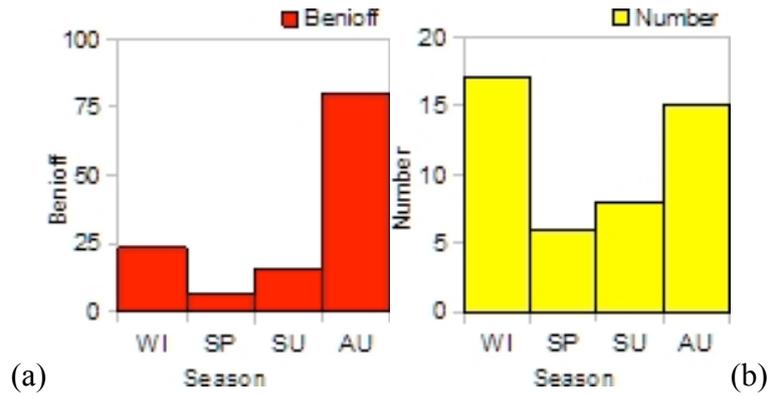

*Fig. 2 – Histograms of Σ (Benioff normalized stress release) and N (number) for the crustal events that occurred in the Circum-Antarctic region (Fig. 1c) in the time interval 1990-2013 with $M_i \geq 6.4$.*

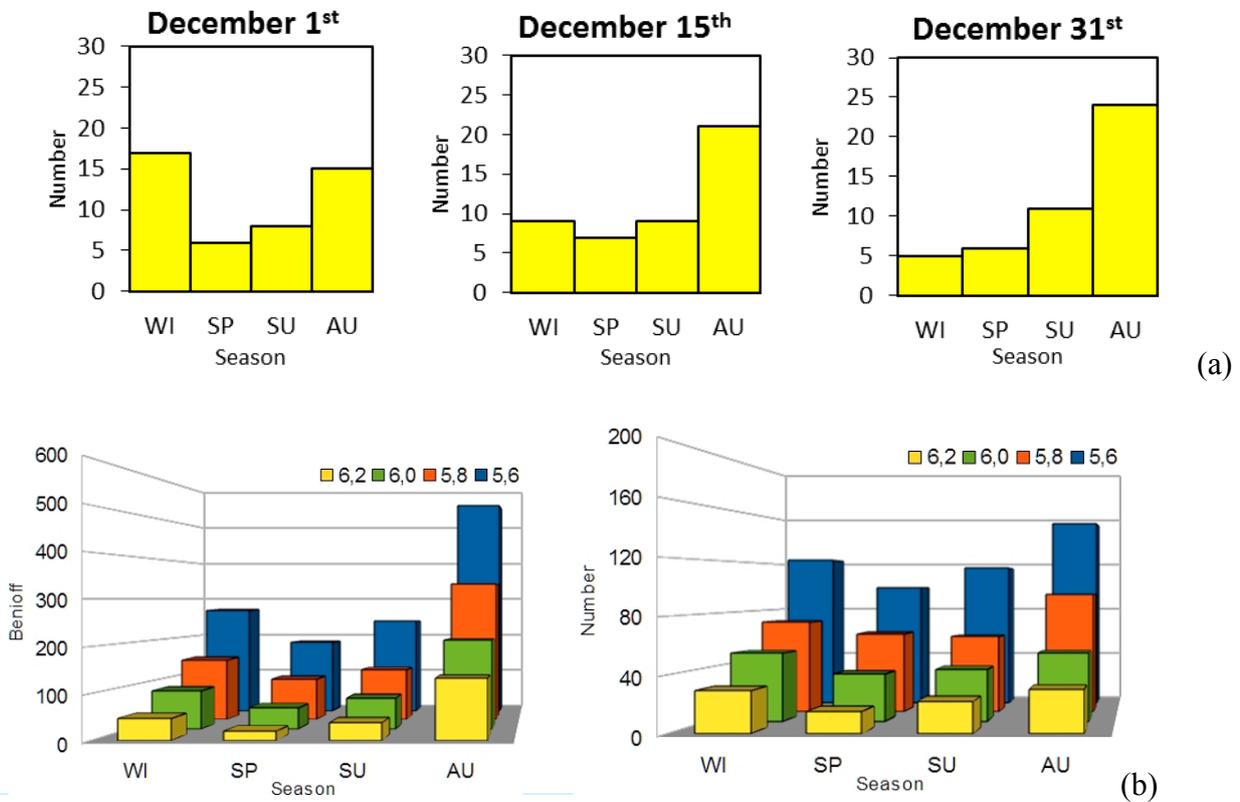

*Fig. 3 – Stability test for the number of crustal events, N, that occurred in the Circum-Antarctic region. In part (a) the three columns correspond to austral summer starting on December 1$^{st}$, 15$^{th}$ and 31$^{st}$, respectively. In part (b) the results are shown, both for N and Σ, for our preferred case (i.e. December 1$^{st}$) considering four magnitude thresholds, M, in the range from 6.2 to 5.6.*

In Fig. 4 as counter-example the results are shown obtained in two extensional regions, where, as one could expect, no seasonal modulation (both N and Σ) due to snowfall is visible. The histograms represent the earthquakes with $M_i \geq 6.1$ recorded in the period 1990-2013, in the Tropical Atlantic Ridge (Fig. 4b), and in the East Pacific Rise (Fig. 4c).

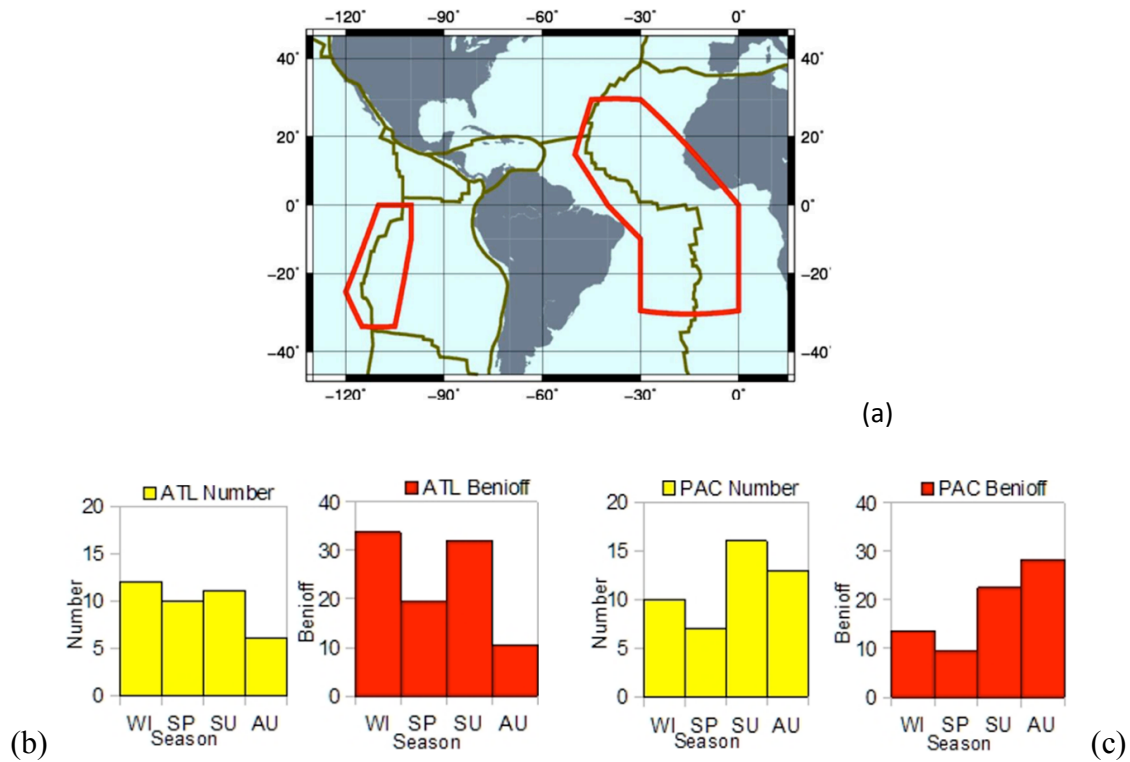

*Fig. 4 – Map (a) and histograms of seismicity (N and Σ) of crustal events for the counter-example regions Tropical Atlantic Ridge (ATL) (b) and East Pacific Rise (PAC) (c). Events occurred in the time interval 1990-2013 with $M_i \geq 6.1$.*

**3. Statistical significance of seasonal modulation in extensional areas**

A quantitative assessment of the seasonal effect, evidenced in section 2, is performed by the classical hypothesis test applied to the number of earthquakes N observed in each season for each study area: Tropical Atlantic Ridge (ATL), East Pacific Rise (PAC) and Circum-Antarctic region.
As a first step, the number N of observed events is compared to a flat distribution, i.e. with no time variation in the number of earthquakes. The flat distribution is the result of computing the mean number of events in each season, using one hundred random catalogues build from the real one by shuffling earthquakes origin time with uniform probability over the year. Only for the real catalogue of the Antarctic region the seasonal distribution differs from average values by more than one standard deviation in every season (Fig. 5a). Differently ATL and PAC distributions, shown in Figs. 5b and 5c respectively, evidence that the real seasonal distribution is entirely (ATL) or partially (PAC) within a standard deviation.

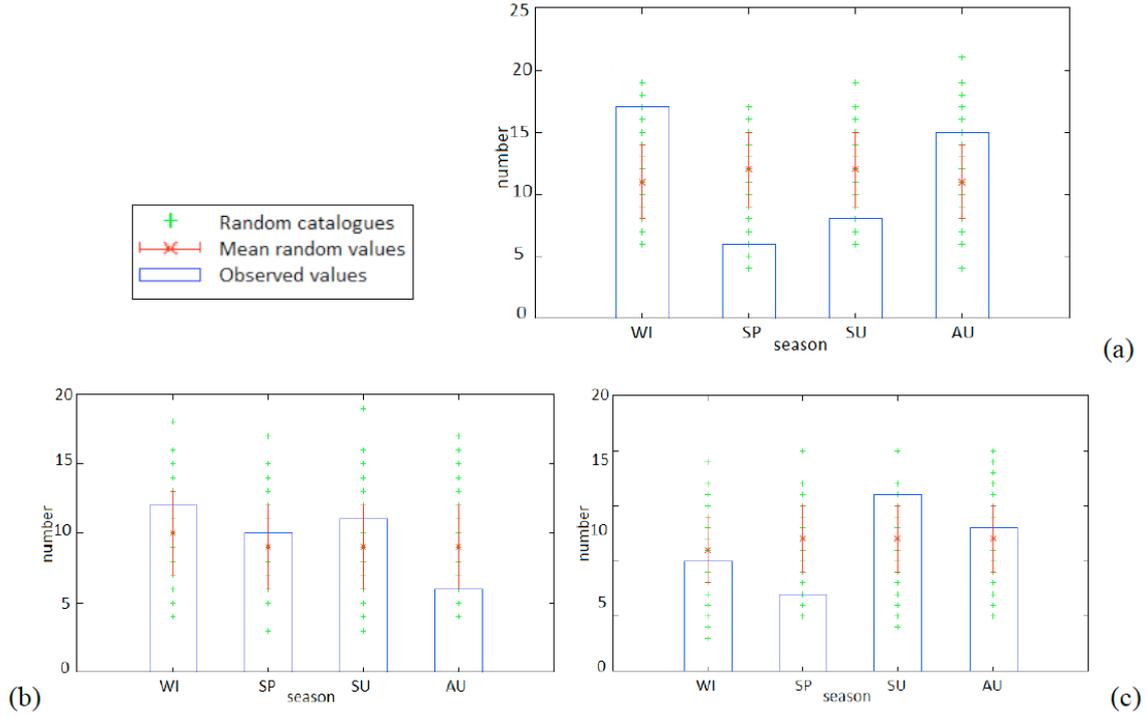

*Fig. 5 – Comparison between seasonal seismicity (N) of randomized catalogues with that of the real catalogue in the Circum-Antarctic Ridge (a) and in the counter-examples regions: Tropical Atlantic Ridge (b) and East Pacific Rise (c). Green crosses represent the results obtained with the hundred randomized catalogues; average values and relative standard deviations are shown in red (crosses and segments, respectively). The histograms represent the distribution of real seismicity (N).*

As second step a parameter is defined for the quantitative assessment of seasonal modulation: the seasonal index

$$SI = \frac{N_h - N_l}{N_{tot}} \quad (2)$$

where $N_h$ represents the seismicity (number of events with M≥M) in warm season (spring-summer) and $N_l$ is the seismicity in cold season (fall-winter). The index $SI$ is bounded in the range $]-1,1[$ and it allows for a quantitative comparison of seismicity in the different periods of year. Similar definition holds replacing $N$ with $\Sigma$.

Obviously, the probability distribution of the seasonal index computed from the randomized catalogues appears to be a normal distribution centred at zero, for the three extensional regions (Fig.6).

If $SI$ is computed for the real catalogues, only in the Antarctic region $SI$ takes a value far from the mean value of normal distribution (see Fig. 6a and Tab. 1). Thus, the hypothesis test shows that a seasonal independent catalogue has got a very small probability to produce the actually observed seasonal index. In particular, this hypothesis can be rejected in the Antarctic region with a confidence level above 99%, considering both N and Σ. In the other extensional regions (ATL and PAC), the hypothesis can be rejected only with confidence level between 49% and 78%. Therefore, in these regions, if there is modulation of the earthquake occurrence, such a modulation is not driven by snow and ice load variations. Therefore, based on the seasonal index metric, *SI,* we cannot discard the hypothesis (with sufficient confidence). We conclude that only in the Antarctic region there is a statistically significant seasonal modulation of seismicity.

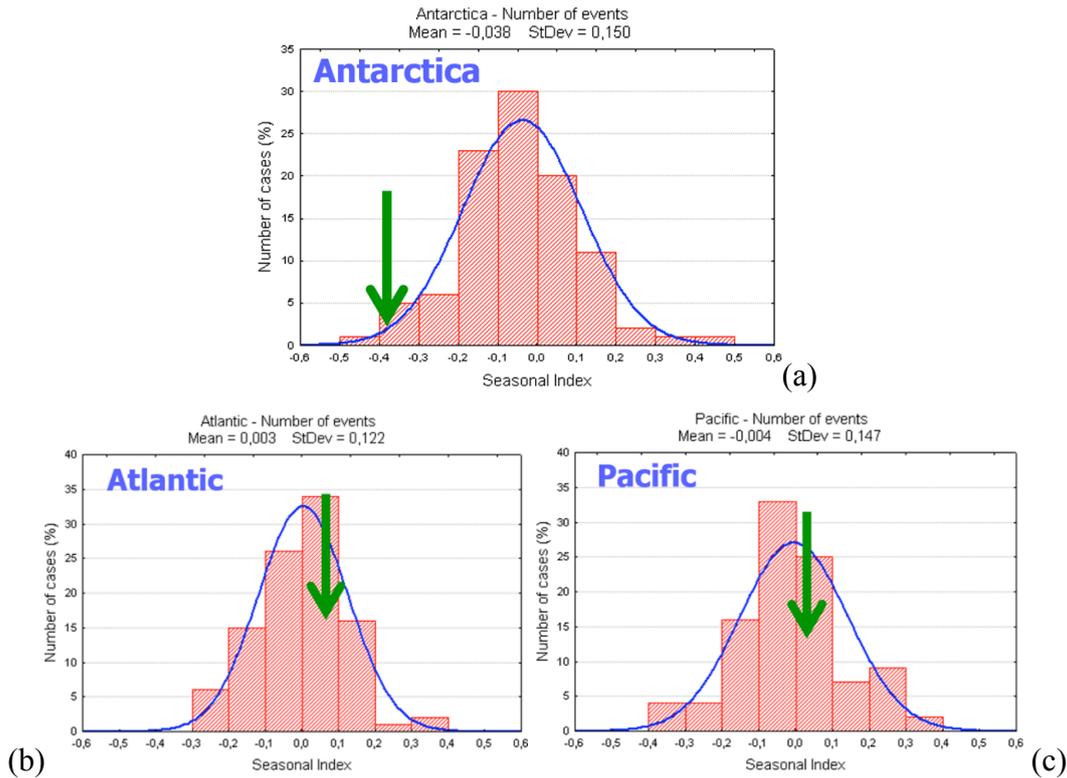

*Fig.6 – Normal distribution of SI calculated from hundred randomized catalogues: a) Circum-Antarctic Ridge, b) Tropical Atlantic Ridge and c) East Pacific Rise. Green arrows show the SIs calculated from the real catalogues.*

| Region | Number of events (N) | | | | | |
|---|---|---|---|---|---|---|
| | Antarctica | | Mid-Atlantic Ridge | | East-Pacific Rise | |
| Season | Real N | N from randomized catalogues | Real N | N from randomized catalogues | Real N | N from randomized catalogues |
| Winter | 17 | 11 ± 3 | 12 | 10 ± 3 | 10 | 11 ± 3 |
| Spring | 6 | 12 ± 3 | 10 | 9 ± 3 | 7 | 11± 3 |
| Summer | 8 | 12 ± 3 | 11 | 9± 3 | 16 | 12± 3 |
| Autumn | 15 | 11 ± 3 | 6 | 9 ± 3 | 13 | 12 ± 3 |

*Tab.1 – Comparison between seasonal seismicity (number of events, N) in real catalogues and averages calculated from the hundred randomized catalogues.*

## 4. Seasonal effect in different tectonic settings

In this section we extend the seasonal seismicity analysis performed in regions already studied by Heki (2003) and Panza et al. (2011), i.e. Japan, Himalaya, Alps and Apennines, to ridge regions, i.e. Circum-Antarctic ridge, Tropical Atlantic Ridge and East Pacific Rise, and several Northern Hemisphere regions, i.e. Caucasus, Zagros, Elburz and Vrancea. In the Southern Hemisphere, the main mountain range (Andes) essentially strikes N-S and comprises several zones, very different in size and climatic conditions, for which the two-dimensional Heki (2003) model is not adequate and warrant a forthcoming special investigation accounting for 3D geometry. Therefore, the quantitative study in the Southern Hemisphere is limited to Antarctica, where the Circum-Antarctic Ridge, as a whole, reacts to the seasonal thaw on the continent (as shown in section 2).

The seismicity analysis follows the scheme described in section 3. The earthquake data sources considered for the Himalaya, Alps and Apennines are described in Panza et al. (2011), updated to using NEIC data (ANSS, 2017). In all other regions we use the global NEIC catalogue (NEIC-USGS, 1989; Shebalin, 1992; ANSS, 2017), except for Vrancea, where ROMPLUS catalogue (Oncescu et al., 1999 and its updates) is used. To increase the sample size, the analysis is updated to December 2016. As in earlier study, aftershocks are removed according to Keilis-Borok et al. (1980). The results for each of the study regions are summarized in table 2 in terms of seasonal index.

The first part of the table lists regions according to their tectonic setting: we observe that shortening regions present positive *SIs*; instead, extensional areas present negative ones. In both cases, the absolute values are greater considering Σ than when N is considered.

The last section of the table contains the results obtained for four counter-example regions, i.e. where snow-ice load is not relevant or where mostly subcrustal earthquakes are involved (Vrancea); events deeper than 100 km do not feel surface ice/snow load, as shown by Heki (2003) computations. In these regions, naturally we observe instability of *SI*: often the sign of *SI* changes with changes in the conventional starting days of seasons.

The cumulative seasonal seismicity distributions, for the considered shortening and extensional tectonic settings and for the four counter-examples regions, reported in Fig. 7, show clearly visible seasonal peaks both in shortening (spring-summer) and in extensional regions (autumn-winter). The seasonal cumulative modulation is quantified by the global *SI*s, reported in table 3.

The snow and ice load competes with tectonic forces in shortening region and antagonizes the earthquake occurrence; instead, snow and ice load collaborate with tectonic forces in extensional regions and facilitates earthquake occurrence.

Furthermore, Fig. 8 shows the cumulative seasonal seismicity distribution of regions with no expected seasonal collaborative or competitive effect of snow and ice load. Both for N and for Σ the patterns are not amenable to any of the cases considered; in particular, they show instability, i.e. high seismicity season near low seismicity season, which is incompatible with accumulation and thaw effects.

| Region | Tectonic regime | Threshold Magnitude, M | Observation Time interval | Number of events, N | SI December 1st | SI December 15th | SI December 31st |
|---|---|---|---|---|---|---|---|
| *Alps* | Shortening | 5.7 | 1850-2016 | 20 | 0.20 (0.22) | 0.40 (0.41) | 0.30 (0.34) |
| *Himalaya* | Shortening | 7.0 | 1850-2016 | 38 | 0.32 (0.51) | 0.21 (0.47) | 0.16 (0.43) |
| *Japan (part with snowfall)* | Shortening | 6.5 | 1904-2016 | 23 | 0.30 (0.46) | 0.22 (0.25) | 0.04 (0.15) |
| *Caucasus* | Shortening | 5.0 | 1965-2016 | 46 | 0.22 (0.38) | 0.13 (0.38) | 0.09 (0.38) |
| *Zagros* | Shortening | 5.8 | 1965-2016 | 42 | 0.19 (0.35) | 0.14 (0.23) | 0.10 (0.02) |
| *Elburz* | Shortening | 5.2 | 1965-2016 | 37 | 0.14 (0.37) | 0.14 (0.37) | 0.08 (0.35) |
| *Antarctica* | Extensional | 6.4 | 1990-2016 | 52 | -0.31 (-0.58) | -0.27 (-0.58) | -0.23 (0.34) |
| *Apennines* | Extensional | 5.7 | 1850-2016 | 45 | -0.02 (-0.28) | 0.07 (-0.07) | 0.11 (-0.03) |
| *Mid-Atlantic Ridge* | Extensional | 6.1 | 1990-2016 | 42 | 0.14 (0.15) | 0.14 (0.10) | 0.14 (0.11) |
| *Pacific rise* | Extensional | 6.1 | 1990-2016 | 48 | -0.04 (-0.14) | 0.08 (0.12) | 0.25 (0.27) |
| *Japan (part without snowfall)* | Shortening | 6.5 | 1904-2016 | 24 | -0.08 (-0.61) | 0.08 (-0.09) | 0.08 (-0.07) |
| *Vrancea* | Extensional | 5.5 | 1930-2016 | 36 | 0.00 (0.01) | -0.06 (-0.20) | -0.06 (-0.26) |

*Table 2 - SI = ($N_h$ - $N_l$)/ $N_{tot}$, where $N_h$ and $N_l$ are the number of events of spring-summer (high temperature) and fall-winter (low temperature) respectively, whereas $N_{tot}$ is the total number of events in the investigated region. SIs computed from the normalized Benioff strain release, Σ, are given in parentheses. In the table results are summarized and provide the SI index for three different starting times of boreal winter: December 1st, 15th and 31st, respectively. For Antarctica, these starting times refer to the austral summer.*

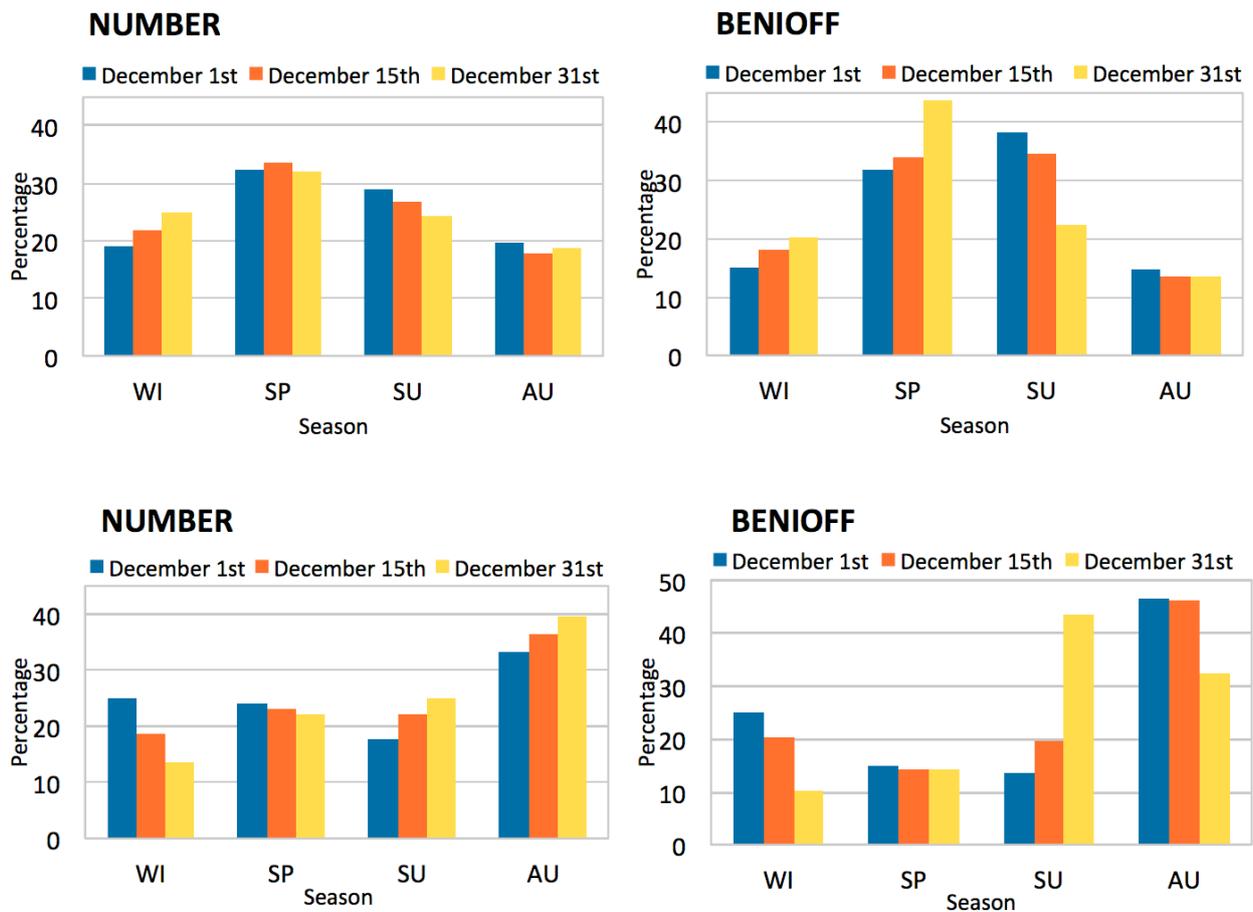

*Fig. 7 - Cumulative histograms of Σ and N expressed as sum of percentages, for the events that occurred in the two tectonic domains considered: shortening regions (upper part), extensional regions (lower part). Shortening regions are Alps, Himalaya, Japan (part A in Fig. 1b), Caucasus, Zagros and Elburz; extensional ones are Apennines and Circum-Antarctic Ridge. The total number of events considered is 206 and 97, respectively, for shortening and extensional regions.*

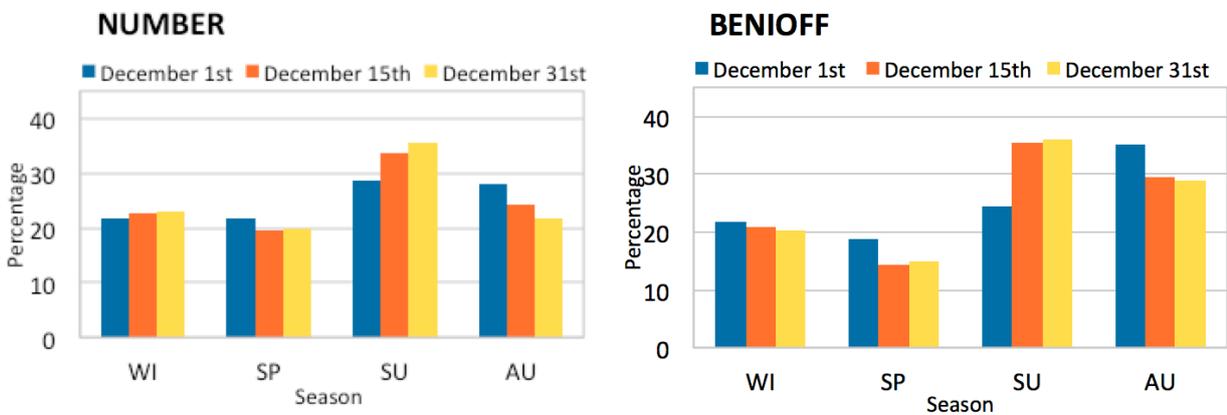

*Fig. 8 - Cumulative histograms of Σ and N expressed as sum of percentages, for the events that occurred in areas with no expected seasonality: Tropical Atlantic Ridge, East Pacific Rise, south Japan (part B in Fig. 1b) and Vrancea (subcrustal events). The total number of events considered is 150.*

| Tectonic setting | Seasonal Index | | |
|---|---|---|---|
| | December 1$^{st}$ | December 15$^{th}$ | December 31$^{st}$ |
| Shortening regions | 0.23 (0.40) | 0.21 (0.37) | 0.13 (0.32) |
| Extensional regions | -0.16 (-0.43) | -0.10 (-0.33) | -0.06 (0.16) |
| Counter- example regions | 0.00 (-0.14) | 0.06 (-0.01) | 0.11 (0.02) |

*Table 3 – Cumulative SIs for the three types of regions considered. Results are summarized, providing SI index for three different starting times of boreal winter: December 1$^{st}$, 15$^{th}$ and 31$^{st}$. SIs computed from Benioff strain release are given in parentheses.*

## 5. Conclusions

The seasonal patterns associated with stress modulation as evidenced by earthquake occurrence have been detected in tectonic regions characterized by snow-ice cover changes in time. The detected seasonal modulation of seismicity well correlates with the dominant tectonic regime: in shortening areas (e.g. Himalaya and Alps) the seismicity is peaking in spring and summer; opposite behaviour – peak in fall and winter – is observed in extensional areas (e.g. Apennines and Antarctica).

Our findings generalize the earlier observations made about short-term (seasonal), long-term (secular) and paleo-seismologic (Mörner, 1995; Panza et al. 2011) modulation of seismicity. In particular, the study confirms that snow and ice thaw may cause crustal deformations that modulate seismicity, with obvious consequences on the validity of seismic hazard maps based on the popular probabilistic methods (Scawthorn, 2008; Panza et al., 2014; Geller et al., 2015). In fact, the seasonal and secular modulation of seismicity is hardly compatible with the hypothesis of Poissonian stationary occurrence of strong earthquakes, at the base of any probabilistic assessment of seismic hazard. The spatio-temporal patterns, which characterise seismicity within homogeneous seismogenic areas associated to a dominating geodynamic process (Peresan, 2017), and the seasonal variability of seismic activity evidenced in this study suggest regularities that cannot be explained by random earthquake occurrence. The tectonic style therefore may play a relevant role in controlling the temporal features of earthquakes occurrence; the coexistence of diametrical tectonic regimes within very narrow distances (e.g. Italy) increases the complexity of the phenomenon.

**Acknowledgements**

We are indebted with R. Mosetti and F. Crisciani for in-depth stimulating discussions and for sharing data and observations. We are also grateful to M. Maybodian for collaboration in the preliminary analysis of Zagros and Elburz regions. This research has been partly developed in the framework of the projects PRIN 2011, Title: "Subduction and exhumation of continental lithosphere: implications on orogenic architecture, environment and climate" (ref. 2010PMKZX7) and PRIN 2015, Title: "The subduction and exhumation of the continental lithosphere: their effects on the structure and evolution of the orogens" (ref. 2015EC9PJ5).